\documentclass[9pt, conference]{IEEEtran}
\IEEEoverridecommandlockouts

\usepackage{cite}
\usepackage{amsmath,amssymb,amsfonts}
\usepackage{algorithmic}
\usepackage{graphicx}
\usepackage{textcomp}
\usepackage{xcolor}
\def\BibTeX{{\rm B\kern-.05em{\sc i\kern-.025em b}\kern-.08em
    T\kern-.1667em\lower.7ex\hbox{E}\kern-.125emX}}

\usepackage{amsmath,amsfonts,bm}

\def\eqref#1{Equation~\ref{#1}}

\def\1{\bm{1}}

\def\rvx{{\mathbf{x}}}

\def\rvz{{\mathbf{z}}}

\DeclareMathAlphabet{\mathsfit}{\encodingdefault}{\sfdefault}{m}{sl}
\SetMathAlphabet{\mathsfit}{bold}{\encodingdefault}{\sfdefault}{bx}{n}

\newcommand{\Ls}{\mathcal{L}}

\usepackage{booktabs} 
\usepackage{multirow} 
\usepackage{hyperref} 
\hypersetup{
    colorlinks=true,
    urlcolor=magenta,
}
\usepackage{pifont} 
\newcommand{\vpm}[2]{$#1 ^{\pm #2}$}  

\begin{document}
\title{Learning Source Disentanglement in Neural Audio Codec
\thanks{This work was funded by the European Union (ERC, HI-Audio, 101052978). Views and opinions expressed are however those of the author(s) only and do not necessarily reflect those of the European Union or the European Research Council. Neither the European Union nor the granting authority can be held responsible for them. This project was provided with computer and storage resources by GENCI at IDRIS thanks to the grant 2024-AD011015054 on the supercomputer Jean Zay's V100 and A100 partition.}
}


  



\author{\IEEEauthorblockN{Xiaoyu Bie\IEEEauthorrefmark{1}, Xubo Liu\IEEEauthorrefmark{2}, Gaël Richard\IEEEauthorrefmark{1}}
\IEEEauthorblockA{\IEEEauthorrefmark{1}LTCI, Télécom Paris, Institut Polytechnique de Paris, France}
\IEEEauthorblockA{\IEEEauthorrefmark{2}CVSSP, University of Surrey, UK}}

\maketitle

\begin{abstract}
Neural audio codecs have significantly advanced audio compression by efficiently converting continuous audio signals into discrete tokens. These codecs preserve high-quality sound and enable sophisticated sound generation through generative models trained on these tokens.  However, existing neural codec models are typically trained on large, undifferentiated audio datasets, neglecting the essential discrepancies between sound domains like speech, music, and environmental sound effects. This oversight complicates data modeling and poses additional challenges to the controllability of sound generation. To tackle these issues, we introduce the Source-Disentangled Neural Audio Codec (SD-Codec),  a novel approach that combines audio coding and source separation. By jointly learning audio resynthesis and separation, SD-Codec explicitly assigns audio signals from different domains to distinct codebooks, sets of discrete representations. Experimental results indicate that SD-Codec not only maintains competitive resynthesis quality but also, supported by the separation results, demonstrates successful disentanglement of different sources in the latent space, thereby enhancing interpretability in audio codec and providing potential finer control over the audio generation process.
\end{abstract}

\section{Introduction}
\label{sec:intro}

Audio codecs have long played a crucial role in audio data storage and transmission. Traditional methods~\cite{dietz2015overview,opus2012} rely on handcraft audio features and signal coding tools. These approaches compress the raw waveform into codec representations and then decompress it back to the original signal, leading to high-quality reconstruction at an adequate bitrate. Training neural networks to transform signals in codecs is known as neural audio codec (NAC). A typical pipeline consists of an encoder and a decoder for signal transformation, as well as a quantizer between them for information coding. Recent advances in NAC by using residual vector quantization (RVQ)~\cite{zeghidour2021soundstream, defossez2022high,kumar2023high} have achieved higher fidelity and lower bitrates compared to traditional methods. Moreover, the ability to reconstruct high quality audio from discrete latent codes has opened the door for realistic audio generation when combined with language models~\cite{dhariwal2020jukebox,lakhotia2021generative,borsos2023audiolm,kreuk2023audiogen} or diffusion models~\cite{san2024discrete}.

Despite the impressive performance of current NAC models, they are typically trained on a mixture of different audio datasets, compressing audio signals from diverse domains into a single latent space. This unified model neglects the fundamental differences between audio from various source domains. For example, the harmonics in human voice differ significantly from those in musical instruments, and the melodies, which are widely present in musics, are rarely seen in ambient sounds. Although, this discrepancy has been acknowledged in the traditional audio coding field with some solutions such as the Unified Speech Audio Codec (USAC)~\cite{neuendorf2009unified}, there is still a lack of investigation into this issue in deep learning based methods. We argue that this may limit the performance in diverse audio environments, especially the explainability of latent features. Addressing this is crucial for future advancements in audio compression and generation.

In this work, we propose the Source Disentangled Neural Audio Codec (SD-Codec), a novel framework that integrates latent source disentanglement into the neural audio codec framework. Given an audio input that may contain multiple sources, we design multiple domain-specific quantizers, each corresponding to audio sources including speech, music, and sound effects (SFX)~\cite{petermann2022cocktail}. SD-Codec learns to disentangle the latent features and distribute features of source domains to different quantizers. The decoder of SD-Codec can reconstruct either a single source track from a specific quantizer, or reconstruct the mixture using the sum of quantized latent features. Furthermore, inspired from~\cite{ginies2024random}, we explore the source disentanglement of SD-Codec by sharing the last several layers of the quantizers. Aiming at a generalized audio codec model, we follow the instructions from~\cite{petermann2022cocktail} and~\cite{pons2024gass} to build mixed training audio clips across different datasets and train our models on a large scale with around $6,000$ hours, as shown in Tab.~\ref{tab:data}. We evaluate our approach on an unseen dataset Divide and Remaster (DnR)~\cite{petermann2022cocktail} and compare it with state-of-the-art methods.

Our key contributions can be summarized as follows:
\begin{itemize}
    \item We introduce SD-Codec, a neural audio codec that can reconstruct audio and separate sources (e.g., speech, music, sound effects) from input audio clips. 
    \item We investigate using a shared RVQ in SD-Codec. The similar performance with and without a shared codebook suggests that RVQ’s shallow layers encode semantic information, while its deeper layers capture local acoustic details.
    \item We train our model on large scale dataset and demonstrate that SD-Codec achieves strong performance both on audio reconstruction and source separation.
\end{itemize}

Code and pretrained models will be open-released, the demo page can be found at: \href{https://xiaoyubie1994.github.io/sdcodec}{https://xiaoyubie1994.github.io/sdcodec}.

\section{Related work}
\label{sec:relate}

\begin{figure*}[!ht]
    \centering
    \includegraphics[width=0.7\textwidth]{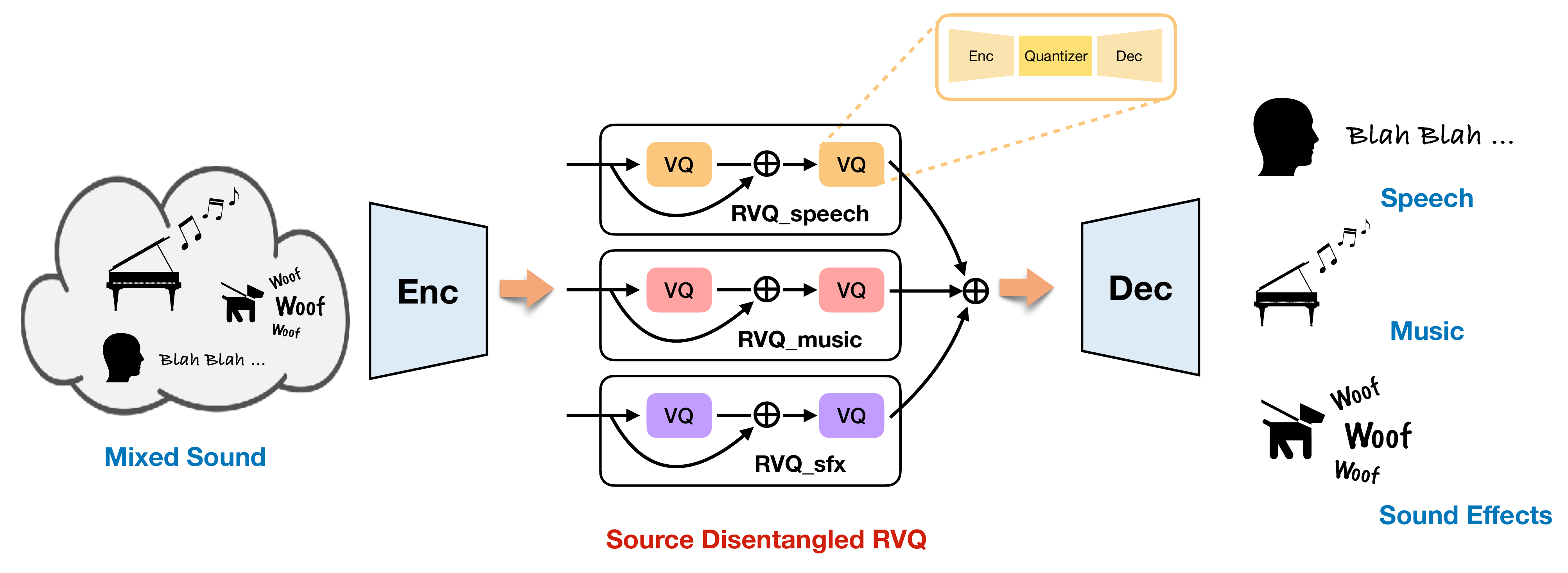}
    \caption{Source Dissentangled Neural Audio Codec (SD-Codec)}
    \label{fig:teaser}
\end{figure*}
\begin{figure}[ht]
    \centering
    \includegraphics[width=0.7\linewidth]{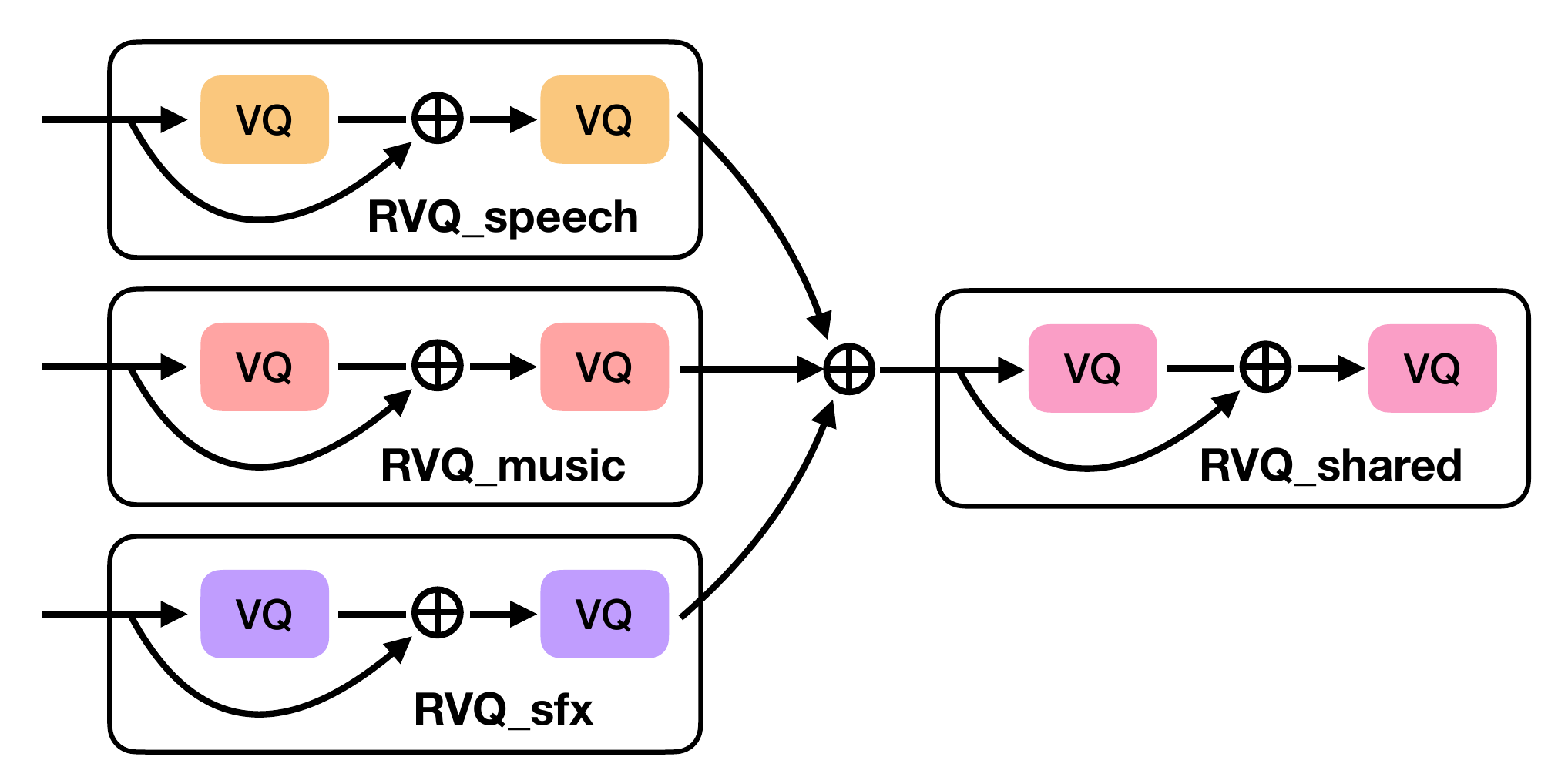}
    \caption{SD-Codec with shared codebooks ($R=4$, $S=2$)}
    \label{fig:shared}
\end{figure}

\subsection{Neural Audio Codec}
\label{sec:relate:codec}
Early approaches for end-to-end neural audio codecs can be traced back to 1990s~\cite{morishima1990speech} and have seen significant advancements with the introduction of deep convolutional neural networks~\cite{kankanahalli2018end}. Concurrently, VQ-VAE~\cite{van2017neural} introduced a learnable quantizer between the encoder and decoder, optimizing them within a probabilistic modeling framework. This approach has proven effective for neural audio codecs operating at low bit rates~\cite{garbacea2019low,kleijn2021generative,yang2024generative}. Subsequently, residual vector quantization and adversarial training were incorporated into VQ-VAE-based neural audio codecs, typical models like SoundStream~\cite{zeghidour2021soundstream} and EnCodec~\cite{defossez2022high} have demonstrated impressive performance in reconstruction quality. More recently, DAC~\cite{kumar2023high} introduced vector factorization and L2-normalization to enhance codebook usage and achieves state-of-the-art on audio reconstruction quality.

\subsection{Latent Feature Disentanglement}
\label{sec:relate:disentangle}
In parallel to improving the quality of reconstruction, some studies dive into learning disentangled latent features for neural audio codec. Multiple encoders were introduced in~\cite{polyak2021speech} to extract phonemes, pitch and speaker identity respectively with different training targets. Later in NaturalSpeech3~\cite{ju2024naturalspeech}, a unified encoder was shown to be sufficient. In addition to disentangling semantic information, Gull~\cite{luo2024gull} learn to disentangle different bandwidth, whereas~\cite{yang2021source, omran2023disentangling} learn to separate noisy or reverberated speech. Our work is most related to~\cite{omran2023disentangling}, but we consider a more general case where human speech, music and environment sound effects are all mixed up, and we do not force the  latent feature to disentangle each audio source into certain channels, but let the quantizer to handle it.

\subsection{Audio Source Separation}
\label{sec:relate:separation}
While supervised speech separation has been widely studied over the decade~\cite{wang2018supervised}, currrent approaches usually compress the audio signals into a latent space and learn a specific separation network to separate different audio sources, such as ConvTasNet~\cite{luo2019conv} and ResUNet~\cite{kong2021decoupling}. Latterly, TDANet~\cite{li2023efficient} achieved great advance by using the feature pyramid and repeating the separation module multiple times. Apart from learning a separation model on a single dataset, recent works, such as GASS~\cite{pons2024gass} and AudioSep~\cite{liu2023separate}, scale up the size of training data, targeting a single model for universal source separation.

\begin{table}[t!]
\caption{Summary of the datasets used for training.}
\vspace{-2mm}
\label{tab:data}
\centering
\resizebox{0.85\linewidth}{!}{
\setlength{\tabcolsep}{1.5mm}
\renewcommand{\arraystretch}{1.2}
{
\begin{tabular}{ccccc}
\toprule
Dataset & Speech & Music & SFX & \# Recordings \\
\midrule
DNS 5~\cite{dubey2023icassp} & \ding{51} & \ding{55} &  \ding{51} & $1,185,771$ \\
MTG-Jamendo~\cite{bogdanov2019mtg}  & \ding{55} & \ding{51} &  \ding{55} & $55,701$\\
MUSAN~\cite{snyder2015musan} & \ding{51} & \ding{51} &  \ding{51} & $1,983$ \\
WHAM!~\cite{wichern2019wham} & \ding{55} & \ding{55} &  \ding{51} & $1,575$ \\
\midrule
Summary & $2,619$h & $3,819$h & $261$h &  $1,245,030$ \\
\bottomrule
\end{tabular}
}
}
\end{table}

\section{Method}
\label{sec:method}

Our model is based on DAC~\cite{kumar2023high},
 designed for $16$ kHz. We describe the details of SD-Codec below.

\begin{table*}[t!]
\caption{Results of zero-shot audio resynthesis and source separation.}
\vspace{-2mm}
\label{tab:main}
\centering
\resizebox{0.70\textwidth}{!}{
\setlength{\tabcolsep}{1.5mm}
\renewcommand{\arraystretch}{1.2}
{
\begin{tabular}{ c | c | cc | cc | cc | cc |}
\toprule
 & & \multicolumn{8}{c|}{Audio Resynthesis} 
 \\
 & & \multicolumn{2}{c|}{Mix} & \multicolumn{2}{c|}{Speech} & \multicolumn{2}{c|}{Music} & \multicolumn{2}{c|}{Sound Effects}  \\
 &  Method & SI-SDR($\uparrow$) & VisQOL($\uparrow$) & SI-SDR($\uparrow$) & VisQOL($\uparrow$) & SI-SDR($\uparrow$) & VisQOL($\uparrow$) & SI-SDR($\uparrow$) & VisQOL($\uparrow$)   \\
\midrule
\multirow{2}{*}{\rotatebox[origin=c]{90}{\textbf{Val}}} & DAC~\cite{kumar2023high} 
& \vpm{4.52}{2.21} & \vpm{4.12}{0.19} & \vpm{7.60}{2.58} & \vpm{4.49}{0.12} & \vpm{5.22}{4.03} & \vpm{4.17}{0.21} & \vpm{0.85}{6.02} & \vpm{3.95}{0.34} \\
& SD-Codec & \vpm{7.02}{2.83} & \vpm{4.28}{0.17} & \vpm{8.33}{3.67} & \vpm{4.45}{0.18} & \vpm{7.72}{4.91} & \vpm{4.03}{0.30} & \vpm{2.32}{6.73} & \vpm{3.96}{0.37} 
 \\
\midrule
\multirow{2}{*}{\rotatebox[origin=c]{90}{\textbf{Test}}} & DAC~\cite{kumar2023high} &
\vpm{4.57}{1.98} & \vpm{4.13}{0.17} & \vpm{7.63}{2.29} & \vpm{4.49}{0.10} & \vpm{5.20}{3.77} & \vpm{4.17}{0.19} & \vpm{1.25}{5.10} & \vpm{3.98}{0.31} \\
& SD-Codec & \vpm{6.98}{2.49} & \vpm{4.29}{0.15} & \vpm{8.28}{3.26} & \vpm{4.44}{0.15} & \vpm{7.65}{4.60} & \vpm{4.03}{0.28} & \vpm{2.54}{5.65} & \vpm{3.98}{0.34} 
  \\
\bottomrule
\end{tabular}
}}
\vspace{2mm}
\resizebox{0.70\textwidth}{!}{
\setlength{\tabcolsep}{1.5mm}
\renewcommand{\arraystretch}{1.2}
{
\begin{tabular}{ c | c |  ccc | ccc | ccc |}
\toprule
 & &  \multicolumn{9}{c|}{Source Separation} \\
 & &  \multicolumn{3}{c|}{Speech} & \multicolumn{3}{c|}{Music} & \multicolumn{3}{c|}{Sound Effects} \\
 &  Method &  SI-SDR($\uparrow$) &  SI-SDRi($\uparrow$) & VisQOL($\uparrow$) & SI-SDR($\uparrow$) &  SI-SDRi($\uparrow$) & VisQOL($\uparrow$) & SI-SDR($\uparrow$) & SI-SDRi($\uparrow$) & VisQOL($\uparrow$)   \\
\midrule
\multirow{2}{*}{\rotatebox[origin=c]{90}{\textbf{Val}}} & TDANet~\cite{li2023efficient} &  \vpm{11.95}{2.97} & \vpm{9.75}{3.16} & \vpm{3.10}{0.46} & \vpm{1.94}{4.34} & \vpm{8.74}{4.05} & \vpm{2.63}{0.58} & \vpm{0.88}{6.16} & \vpm{8.41}{4.29} & \vpm{2.33}{0.78} \\ 
& SD-Codec &  \vpm{11.26}{3.35} & \vpm{9.07}{3.13} & \vpm{3.41}{0.48} & \vpm{1.73}{4.23} & \vpm{8.53}{3.45} & \vpm{2.84}{0.65} & \vpm{0.91}{5.23} & \vpm{8.45}{3.88} & \vpm{2.44}{0.79} \\
\midrule
\multirow{2}{*}{\rotatebox[origin=c]{90}{\textbf{Test}}} & TDANet~\cite{li2023efficient} & \vpm{11.86}{2.66} & \vpm{9.91}{2.94} & \vpm{3.21}{0.39} & \vpm{2.12}{3.81} & \vpm{8.83}{3.68} & \vpm{2.74}{0.50} & \vpm{1.87}{4.79} & \vpm{8.62}{3.45} & \vpm{2.49}{0.72}  \\ 
& SD-Codec & \vpm{11.31}{2.98} & \vpm{9.36}{2.80} & \vpm{3.49}{0.40} & \vpm{1.85}{3.68} & \vpm{8.57}{3.04} & \vpm{2.96}{0.56} & \vpm{1.77}{4.08} & \vpm{8.52}{3.23} & \vpm{2.64}{0.72}  \\
\bottomrule
\end{tabular}
}
}

\end{table*}

\subsection{Model Architecture}
\label{sec:method:model}
The architecture of our model is shown in Fig.~\ref{fig:teaser}. The input is a single-channel audio waveform $\rvx \in \mathbb{R}^T$, this audio signal can either be a single source from speech, music, sound effects, or any combination of them. The encoder (\textbf{Enc}) consists of an input 1D convolution, followed by several residual convolution blocks, and ends up by another 1D convolution. The latent features obtained from the encoder are $\rvz= Enc(\rvx) \in \mathbb{R}^{F \times D}$, where $F$ is the number of frames and $D$ is the latent feature dimension. The decoder (\textbf{Dec}) is a mirror structure of the encoder. We use the same hyper-parameters as in DAC-$16$kHz~\cite{kumar2023high} for the architecture design, the latent features thus have a dimension of $1,024$ and have $50$ latent frames per second.

For quantization, We use the residual vector quantizer (\textbf{RVQ}). Each VQ layer consists of a projection layer to compress the latent features to lower dimensional space ($d=8$). The compressed latent features will be assigned to a nearest codebook centroid based on Euclidean distance after L2-normalization, then the quantized features will be projected back to its orignal dimension $D$. Different from DAC, SD-Codec have three domain-specific RVQs including $\text{RVQ}_{\text{speech}}$, $\text{RVQ}_{\text{music}}$ and $\text{RVQ}_{\text{sfx}}$. Each RVQ has $R$ residual VQ layers and learns domain-disentangled latent features, which can either be used to reconstruct the audio for a specific source domain, or can be added up to reconstruct the mixture. We further design a variant of SD-Codec where all the domain-specific RVQs share the last $S$ VQ layers, i.e. a $\text{RVQ}_{\text{shared}}$ is concatenated after domain-specific RVQs. Fig.~\ref{fig:shared} gives an example of $R=4$ and $S=2$. For example, if we assume the codebook size in each layer is $2^K$, then the bits allocated for each frame is $R \times K$. In practice, we use $R=12$ and $K=10$ to have a bitrate of $6,000$ bits per second ($6$ kbps), which is the same as DAC-16kHz for single source resynthesis.

\subsection{Source Disentanglement and Resynthesis}
\label{sec:method:disentangle}

We consider a mixture speech, music and sound effects, the latent features from the encoder can be written as:
\begin{align}
    & \rvx_{\text{mix}} = \rvx_{\text{speech}} + \rvx_{\text{music}} + \rvx_{\text{sfx}} \\
    & \rvz = Enc(\rvx_{\text{mix}})
\end{align}

Then, each source-aware RVQ learns the specific features related to its source, formally:
\begin{align}
    & \hat{\rvz}_s^{(0)} = 0, \\
    & \hat{\rvz}_s^{(i+1)} = VQ_{s}^{(i+1)}(\rvz - \sum_{r=0}^{i} \hat{\rvz}_s^{(r)}),
\end{align}
where $s \in \{\text{speech}, \text{music}, \text{sfx}\}$ and $VQ_{s}^{(i+1)}$ represents the $(i+1)$th quantization layer for the source $s$.

To reconstruct a single source such as speech:
\begin{align}
    & \hat{\rvz}_{\text{speech}} = \sum_{i=1}^{R} \rvz_{\text{speech}}^{(i)}, \\
    & \hat{\rvx}_{\text{speech}} = Dec(\hat{\rvz}_{\text{speech}}).
\end{align}

To reconstruct the mixture, we assume the linear additivity in the latent space:
\begin{align}
    & \hat{\rvz}_{\text{mix}} = \hat{\rvz}_{\text{speech}} + \hat{\rvz}_{\text{music}} + \hat{\rvz}_{\text{sfx}}, \\
    & \hat{\rvx}_{\text{mix}} = Dec(\hat{\rvz}_{\text{mix}}).
\end{align}
While this shows the case of three-source mixture, it is also valid for the mixture of two sources. Here we need three times more bitrate for $3$-source mixture than single source. To be noted that, if we share all the codebooks, i.e. $S=R$, then the SD-Codec will be the same as the DAC model, whereas if there is no codebook shared, then SD-Codec will consist of three RVQ in parallel. 

\subsection{Training Objective}
\label{sec:method:train}
Since SD-Codec is able to reconstruct four audio (three single source and one mixture), the final loss is:
\begin{equation}
    \Ls_{tot} = \sum_{s} \Ls_{g} (\hat{\rvx}_{s}, \rvx_{s}),
\end{equation}
where $s \in \{ \text{mix}, \text{speech}, \text{music}, \text{sfx} \}$. The $\Ls_g$ is the same generator loss used in DAC~\cite{kumar2023high}, which consists of the multi-scale mel-spectrograms loss, the feature matching loss, the codebook loss and the commitment loss. Same as DAC, we also use a multi-period waveform discriminator (MPD) and a complex STFT discriminator to compute the feature matching loss and the adversarial loss. Following~\cite{omran2023disentangling}, we randomly shuffle the latent feature combination in the batch to enhance the feature disentanglement. Moreover, to balance the model performance on resynthesis and separation, we randomly select the number of audio tracks used to make up the mix. In our experiments, the probabilities of using $1$, $2$, and $3$ tracks are $0.6$, $0.2$, and $0.2$, respectively.

\section{Experiments}
\label{sec:exp}

\subsection{Dataset}
\label{sec:exp:data}

We trained our models on a large dataset across speech, music, and sound effects. As summarized in Tab.~\ref{tab:data}, we used the clean speech corpus and noise corpus from DNS Challenge 5~\cite{dubey2023icassp}, MTG-Jamendo~\cite{bogdanov2019mtg}, WHAM!~\cite{wichern2019wham} and MUSAN~\cite{snyder2015musan}. In general, the training dataset is compiled of $2,619$ hours of speech data, $3,819$ hours of music data, and $261$ hours of sound effects data. To evaluate the generalized performance of our models, we used the validation set and test set from DnR~\cite{petermann2022cocktail}. Since the DnR dataset is composed of artificial mixtures using speech, music, and sound effects which are different from our training sources, the validation and test can be considered as zero-shot, or out-of-distribution evaluation. We trimmed the leading and trailing silence in the mixture audio, and clipped the remaining audio into segments with a length either of $5$-second (validation), or of $10$-second (test). We also removed those segments where the individual tracks in the mix were too short (less than 50\% of the mixture). Finally, the evaluation dataset consisted of $2,853$ mixtures ($\approx$ $4$ h) from the validation and $1,840$ mixtures ($\approx$ $5$ h) from the test.

For training, we followed the pre-processing in DnR data composition~\footnote{\href{https://github.com/darius522/dnr-utils/}{https://github.com/darius522/dnr-utils/}}. We first normalized each source to a specific loudness units full-scale (LUFS)~\cite{grimm2010lufs}. For speech, music and sound effects, the target LUFS values were $-17$, $-24$ and $-21$ dB respectively, with additional random perturbation uniformly sampled from $\pm$2. Then we normalized the peak value to $-0.5$ dB if any of the audio sources had a peak value larger than this threshold. Finally, we mixed up all the individual audio sources and normalize the mixture to $-27$ LUFS with random perturbation.

\begin{table*}[t!]
\caption{Results of ablation study on our proposed codec.}
\vspace{-2mm}
\label{tab:abla}
\centering
\resizebox{0.7\textwidth}{!}{
\setlength{\tabcolsep}{1.5mm}{
\begin{tabular}{ c | c | cccc | ccc |}
\toprule
 & & \multicolumn{4}{c|}{Re-synthesis (SI-SDR $\uparrow$)} & \multicolumn{3}{c|}{Source Separation (SI-SDR $\uparrow$)} \\
& Method & Mix & Speech & Music & Sfx & Speech & Music & Sfx \\
\midrule
\multirow{5}{*}{\rotatebox[origin=c]{90}{\textbf{Val}}} & SD-Codec & \vpm{7.02}{2.83} & \vpm{8.33}{3.67} & \vpm{7.72}{4.91} & \vpm{2.32}{6.73} & \vpm{11.26}{3.35} & \vpm{1.73}{4.23} & \vpm{0.91}{5.23}   \\
& + shared codebook (S=4) & \vpm{7.19}{2.79} & \vpm{8.65}{3.58} & \vpm{7.74}{4.81} & \vpm{2.44}{6.61} & \vpm{11.18}{3.31} & \vpm{1.60}{4.28} & \vpm{0.66}{5.20} \\
& + shared codebook (S=8) & \vpm{7.14}{2.76} & \vpm{8.60}{3.55} & \vpm{7.78}{4.82} & \vpm{2.47}{6.70} & \vpm{11.13}{3.25} & \vpm{1.67}{4.04} & \vpm{0.67}{5.20} \\
& + separation enhance  & \vpm{7.21}{2.82} & \vpm{7.95}{3.86} & \vpm{6.84}{4.84} & \vpm{1.02}{6.55} & \vpm{11.78}{3.27} & \vpm{2.33}{4.14} & \vpm{1.44}{5.32} \\
& + initization from DAC & \vpm{-5.20}{2.86} & \vpm{-1.64}{3.69} & \vpm{-3.43}{5.34} & \vpm{-14.08}{11.61} & \vpm{10.28}{3.15} & \vpm{0.70}{4.27} & \vpm{-0.12}{5.06}  \\
\midrule
\multirow{5}{*}{\rotatebox[origin=c]{90}{\textbf{Test}}} & SD-Codec & \vpm{6.98}{2.49} & \vpm{8.28}{3.26} & \vpm{7.65}{4.60} & \vpm{2.54}{5.65} & \vpm{11.31}{2.98} & \vpm{1.85}{3.68} & \vpm{1.77}{4.08}  \\
& + shared codebook (S=4)  & \vpm{7.15}{2.46} & \vpm{8.60}{3.18} & \vpm{7.67}{4.51} & \vpm{2.68}{5.56} & \vpm{11.21}{3.00} & \vpm{1.71}{3.73} & \vpm{1.52}{3.99} \\
& + shared codebook (S=8) & \vpm{7.11}{2.43} & \vpm{8.57}{3.15} & \vpm{7.70}{4.54} & \vpm{2.69}{5.52} & \vpm{11.18}{2.90} & \vpm{1.79}{3.53} & \vpm{1.54}{4.11}  \\
& + separation enhance & \vpm{7.17}{2.48} & \vpm{7.91}{3.40} & \vpm{6.77}{4.56} & \vpm{1.29}{5.53} & \vpm{11.83}{2.91} & \vpm{2.46}{3.61} & \vpm{2.27}{4.23}   \\
& + initization from DAC & \vpm{-5.02}{2.47} & \vpm{-1.54}{3.12} & \vpm{-3.15}{4.83} & \vpm{-12.22}{9.49} & \vpm{10.34}{2.81} & \vpm{0.87}{3.68} & \vpm{0.78}{3.95}  \\
\bottomrule
\end{tabular}
}
}
\end{table*}

\subsection{Evaluation}
\label{sec:exp:eval}
For evaluating the performance on audio resynthesis and source separation tasks, we both use Scale-Invariant Signal-to-Distortion Ratio (SI-SDR)~\cite{le2019sdr} for signal-level audio quality evaluation and ViSQOL~\cite{hines2015visqol} for perceptual audio quality assessment. Since each source has been normalized to different LUFS, we also report SI-SDR improvement (SI-SDRi) in source separation.

\subsection{Training}
\label{sec:exp:train}
We train all models for $400,000$ iterations with the Adam optimizer with a batch size of $64$ examples of $2$ seconds each, a learning rate of $1\cdot 10^{-4}$, $\beta_1=0.8$, and $\beta_2=0.99$. We linearly warm up the learning rate for the first $10,000$ iterations and then apply an exponential scheduler with $\gamma=0.999996$. All the models are trained using $8$ NVIDIA A100 GPUs.

\subsection{Results}
\label{sec:exp:ret}
To evaluate the performance of SD-Codec on audio resynthesis and source separation, we compare it with DAC~\cite{kumar2023high} and TDANet~\cite{li2023efficient}, the state-of-the-art models on audio codec model and source separation.~\footnote{While TDANet was designed for speech separation, it also showed state-of-the-art performance in general source separation in~\cite{pons2024gass}} For fair comparison, we use their official implementations~\footnote{\href{https://github.com/descriptinc/descript-audio-codec}{https://github.com/descriptinc/descript-audio-codec}}~\footnote{\href{https://github.com/JusperLee/TDANet}{https://github.com/JusperLee/TDANet}} and re-train their models on the same training data corpus as SD-Codec. The results are shown in Tab.~\ref{tab:main}. In general, we observe that all the models achieve decent performance on zero-shot evaluation with different data corpus and different data lengths, which confirms the benefit of generalization from scale-up training. 

In audio resynthesis, since SD-Codec has triple bitrates than DAC on mixed audio, there is no surprise that SD-Codec achieves better reconstruction quality. When it comes to single source resynthesis, SD-Codec has the same bitrate as DAC. The reconstructed audio clips from SD-Codec have higher SI-SDR of around $1$ dB than those from DAC, and achieve comparable VisQOL values.

In source separation experiments, we do not use the direct output of the decoder. Instead, we use a mask augmentation method where the decoder output is used to compute a magnitude mask. This mask is then applied to the magnitude of the mixture input, and each audio track is reconstructed using the noisy phase from the mixture. We observed that this operation will improve the SI-SDR, and has slight positive impact on VisQOL values. Compared to TDANet, although SD-Codec is not specifically designed for source separation and experiences quantization loss, it achieves comparable results to TDANet in terms of SI-SDR, SI-SDRi, and VisQOL.

From the above results in Tab.~\ref{tab:main}, we can conclude that our proposed SD-Codec, while explicitly assigning latent features from different sources to different codebooks, can be applied both on audio resynthesis and source separation, and achieve promising results. 

\begin{figure}[th]
    \centering
    \includegraphics[width=0.9\linewidth]{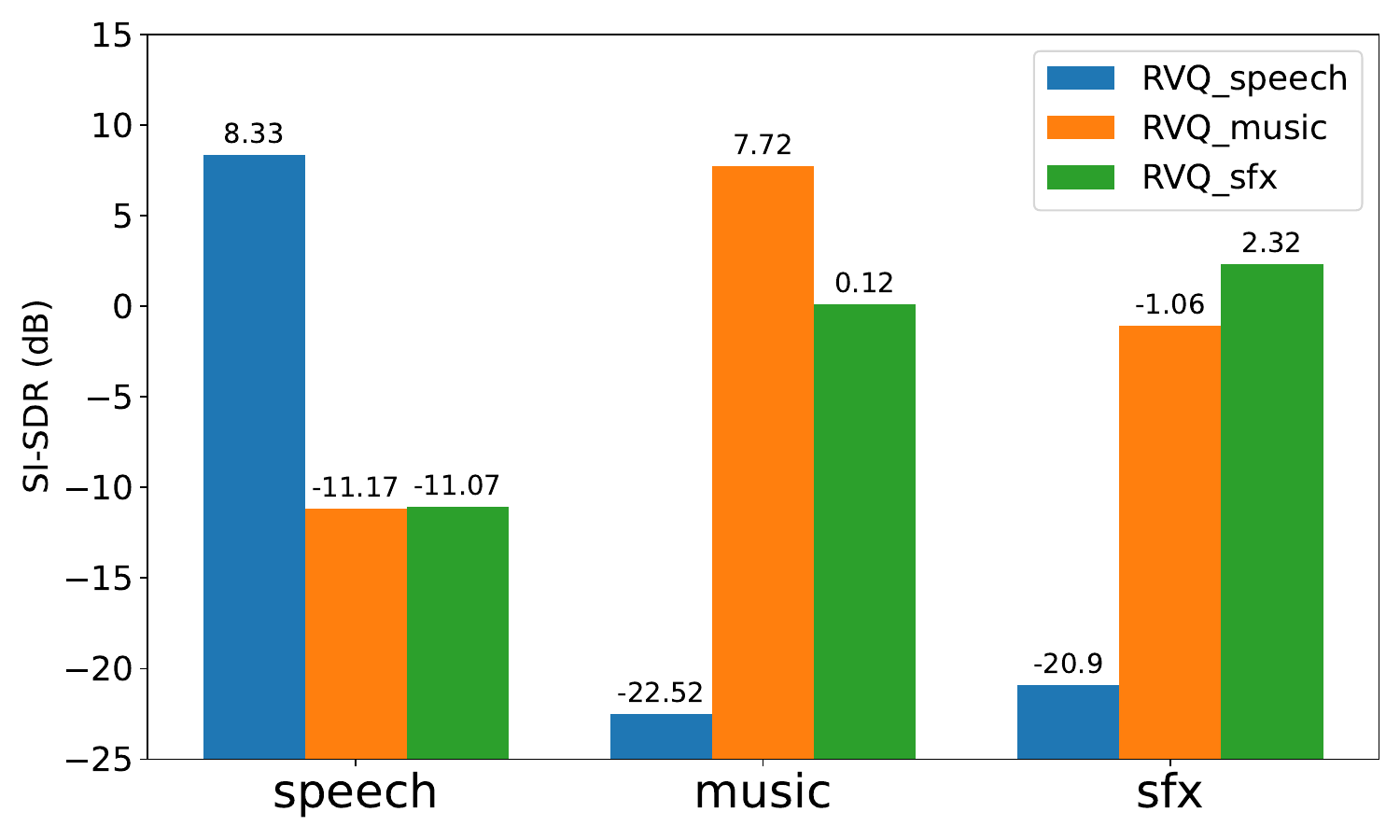}
    \caption{Single source audio resynthesis using different RVQ modules.}
    \label{fig:disentangle}
\end{figure}

To investigate whether SD-Codec learns a domain-specific quantizer, we present the results of reconstructing different audio sources with various RVQ configurations in Fig.~\ref{fig:disentangle}. The results clearly demonstrate that only the corresponding RVQ module can reconstruct the audio signal with high quality.

\subsection{Ablation Study}
\label{sec:exp:abla}

In ablation experiments shown in Tab.~\ref{tab:abla}, we mainly studied the changes brought about by different designs of SD-Codec:

\begin{itemize}
    \item \textit{Shared codebook}: As shown in Fig.~\ref{fig:shared}, we share the last layers for each source. We observe that by either sharing the last $4$ layers or the last $8$ layers, SD-Codec can maintain similar performance. This shows that in RVQ modules, the shallow layers encode the majority of source-aware information, while the deeper layers encode the local acoustic details.  
    \item \textit{Separation enhance}: When training the original SD-Codec, we randomly used $1$, $2$, or $3$ tracks as input with probabilities of $0.6$, $0.2$, and $0.2$, respectively. We also experimented with different probabilities of $0.2$, $0.2$, and $0.6$, which increased the likelihood of the model encountering mixtures of 3 tracks, thereby enhancing its ability to learn separation. As shown in the ablation results, this will bring higher reconstruction quality on the mixture and higher separation quality, with the expense of losing the quality of individual signal reconstruction.

    \item \textit{Initization from DAC}: We also studied the potential to leverage from pre-trained DAC on SD-Codec training. However, due to the huge difference in objective, i.e. learning a unified latent space for all audio sources and learning source disentangled latent space, SD-Codec suffers from the DAC pretraining.
\end{itemize}

\vspace{1em}
\section{Conclusion}
\label{sec:conclustion}
In this work, we designed a source-aware audio codec, SD-Codec. By joint learning audio coding and source separation, SD-Codec obtain excellent results, comparable to the state-of-the-art both on audio reconstruction and separation, with a unified model. Meanwhile, SD-Codec explicitly assigns latent features from different audio domain to different codebook, which brings more explainability and interpretability on the latent features. Furthermore, due to the source disentanglement in the latent space, future audio generation models can have more fine-grained manipulations on the generated audio, paving the way to the research of controllable audio generation.

\newpage
\bibliographystyle{IEEEtran}
\bibliography{bibtex/refs_short,bibtex/bibliography}

\end{document}